\documentclass[]{interact}

\usepackage{epstopdf}
\usepackage[caption=false]{subfig}

\usepackage[numbers,sort&compress,merge]{natbib}
\bibpunct[, ]{[}{]}{,}{n}{,}{,}

\theoremstyle{plain}

\theoremstyle{definition}

\theoremstyle{remark}

\usepackage{amsmath}
\usepackage{amssymb}

\usepackage{comment}

	\newcommand{\be}{\begin{equation}}
		\newcommand{\ee}{\end{equation}}
	\newcommand{\bea}{\begin{eqnarray}}
		\newcommand{\eea}{\end{eqnarray}}
	\newcommand{\ba}{\begin{array}}
		\newcommand{\ea}{\end{array}}

	\newcommand{\bc}{\begin{center}}
		\newcommand{\ec}{\end{center}}	

	\newcommand{\braket}[2]{\left< #1 \vphantom{#2} \right|
		\left. #2 \vphantom{#1} \right>} 
	\let\baraccent=\= 
	\renewcommand{\=}[1]{\stackrel{#1}{=}} 

\begin{document}


\title{Universal information-entropic inequalities for Franck-Condon factors of diatomic molecules.}

\author{
	\name{
		V.~I. Man'ko \textsuperscript{1,2} and Z. Seilov\textsuperscript{2}\thanks{Email: zhanat.seilov@phystech.edu}
	}
	\affil{
		\textsuperscript{1}Lebedev Physical Institute of Russian Academy of Science, 53 Leninskiy Prospekt, Moscow, Russian Federation, 119991;
		\\ \textsuperscript{2}Moscow Institute of Physics and Technology, 9 Institutskiy per., Dolgoprudny, Moscow Region, Russian Federation, 141700
	}
}

\maketitle

\begin{abstract}
New entropic inequalities are obtained for Franck-Condon factors (FCF-s) of diatomic molecules in adiabatic harmonic oscillator approximation. The probability distributions associated with FCF-s of diatomic molecules are presented in the form of joint-probability distributions of either two or three random variables. In view of this representation information inequalities corresponding to subadditivity condition is found and in case of harmonic oscillator approximation is obtained in explicit form of new inequalities for Hermite polynomials.
Physical meaning of obtained entropic-information inequalities as characteristics of hidden correlations in vibronic spectra of the molecules is studied.
\end{abstract}

\begin{abbreviations}
	FCF - Franck-Condon factor, FCI -- Franck-Condon Integrals 
\end{abbreviations}	


\section{Introduction}

The theory of vibronic spectra of molecules was studied during many decades since publications of \cite{Herz50, Cond26}. The structure of the spectra is determined by the FCF-s of the polyatomic molecules. In harmonic approximation these factors are given in terms of overlap integrals of the wave functions of the oscillator's energy states corresponding to nuclear vibration.  Explicit expressions for FCF-s of polyatomic molecules were obtained and analyzed in terms of Hermite polynomials of several variables in \cite{sharp, Dok75, *Dok77, *Dok79} see also \cite{Dok1, Dok2, Dok3, Zheb16, Dush37}. The vibronic spectra are considered in adiabatic approximation where the intensity of the vibronic structure of electronic lines is expressed in terms of FCF-s. The electronic transitions in polyatomic molecules and vibronic structure  of the electronic lines depend on potential well describing the vibration of nuclei in the molecules. Authors of \cite{gmerek} developed a program which computes the FC integrals of multidimensional, harmonic oscillators based on formula given in \cite{Dok75,Dok77} and study  fluorescence emission spectra of the 2-tolunitrile dimer and the 2-tolunitrile water cluster. In \cite{islmpour} were developed the calculations of the vibronic structure in electronic absorption and fluorescence spectra of a large polyatomic molecule based on time- and frequency-domain approaches beyond the Condon approximation.
In \cite{yang2015} were analyzed similarities between indole and 3-methylindole with density functional theory (DFT) and its time-dependent extension (TD-DFT) and were simulated vibrationally resolved $^{1}L_b \leftrightarrow S_0$ electronic spectra  within the Franck-Condon approximation and included the Herzberg-Teller (HT) and Duschinsky effects. In work \cite{tran2018} were explored limitations associated with the fitting a quadratic Hamiltonian to vibronic levels of a Jahn-Teller system.
Contemporary employment of this theory in boson sampling framework is studied in \cite{Huh12, huh15, huh17, huh18}.

The theory of the FCF-s in framework of tomographic probabilities description of quantum states was developed for harmonic approximation in [8]. In this approximation as it was known in \cite{Dok75, *Dok77} the factors are given in terms of multivariable Hermite polynomials. For diatomic molecules these factors are expressed in terms of Hermite polynomials of two variables. The factors $P_n^m=|<n_i|m_f>|^2$ represent the probability distribution for vibronic transitions from initial vibronic state $|n,i>$ of the molecule to final vibronic state $|m,f>$. Development of analytical approach for computing FCIs of harmonic oscillators with arbitrary dimensions in which the mode‐mixing Duschinsky effect is taken into account was studied in \cite{chang} and was applied to study the photoelectron spectroscopy of vinyl alcohol and ovalene (C32H14). In work \cite{meier, osche} were discussed few different approaches for calculating FCF-s beyond the harmonic approximation taking into account Duschinsky effect. In reviews \cite{borrelli, bloino} were discussed different algorithms and methodologies for the calculation of FCIs and different methodologies available for the simulation of vibrational and vibrationally resolved electronic spectra of medium-to-large molecules. As was pointed out in \cite{Huh12} properties of FCF-s known in theory of polyatomic molecules can be applied in quantum information processing.

The aim of our work is to obtain new inequalities for the FCF-s for diatomic molecules. The approach can be extended to the case of polyatomic molecules. These inequalities reflect the presence of hidden correlations which can exist in the vibronic structure of the electronic lines. These correlations are studied in framework of recent consideration [10] of the quantum correlations in quantum information approach based on tomographic probability representation of quantum states [11,12]. As it was found in [13,14,15], the states of noncomposite system can be mapped onto the states of composite ones both for the case of classical system (probability representation) and quantum system (density matrix representation). Due to this one can apply the formalism of von Neumann, Shannon, Tsallis and Renyi entropy to the probability distributions and density matrices. In our work we apply the approach for entropic-information known for joint probability distributions and density matrices  to the FCF-s. Some aspects of this approach were mentioned in \cite{iop18}.

\section{Correlations in vibronic repetitions in electron line intensities for harmonic approximation}

In our research we use probabilistic approach to FCF-s, representing them as probabilities and applying Shannon entropy properties. 

Shannon entropy of a system $A$ associated with the discrete probability distribution $p_i$  is defined as follows 

$ H(A)\equiv H_A(p_i)=-\sum_{i}p_i \log {p_i} $

The property of subadditivity of Shannon entropy for composite system $H(AB)$ consisting of two subsystems -- $A$ and $B$ is expressed a
\be \label{sub}
H(A)+H(B) \geq H(AB),
\ee

For composite system $ AB $ with joint probability distribution $ p(x_1, x_2) $ the marginal probability distributions is known

\be \mathcal{P}(x_1)=\sum_{x_2=1}^{X_2}p(x_1,x_2)=\sum_{x_2=1}^{X_2}f(y(x_1,x_2))
=\sum_{x_2=1}^{X_2}f(x_1+(x_2-1)X_1) \label{marg1}\ee
and

\be \Pi(x_2)=\sum_{x_1=1}^{X_1}p(x_1,x_2)=\sum_{x_1=1}^{X_1}f(y(x_1,x_2))=
\sum_{x_1=1}^{X_1}f(x_1+(x_2-1)X_1), \label{marg2}\ee
 where $1\leq x_1 \leq X_1$, $~1\leq x_2 \leq X_2$

In our work \cite{Sei17} we introduced the approach of dividing one system into few subsystems, i.e. representing variable $n$ in terms of few variables $n(x_1,x_2, ..)$.	The bijective map for the set of integer numbers $y$ and the set of
$n$ integer variables $x_i$ reads
\be
y=y(x_1,x_2,\ldots,x_n)=x_1+\sum_{k=2}^{n}(x_k-1)\prod_{j=1}^{k-1}X_j, \qquad
1\leq x_i \leq X_i,\quad i\in[1,n].        \label{su}\ee
The function $x_k(y)$ is
\be x_k(y)-1=\frac{y-(x_1+\sum_{i=2}^{k}(x_i-1)\prod_{j=1}^{i-1}X_j)}{\prod_{j=1}^{k}X_j}\mod{X_k},
\qquad k=1,\ldots,n, \quad 1\leq y \leq N. \label{us}
\ee

Subadditivity condition (\ref{sub}) can be rewritten as follows:
\be     \label{sub2}
-\sum_{x_1=1}^{X_1}\mathcal{P}(x_1)\log\mathcal{P}(x_1)
-\sum_{x_2=1}^{X_2}\Pi(x_2)\log\Pi(x_2) \geq
-\sum_{x_1=1}^{X_1}\sum_{x_2=1}^{X_2} p(x_1,x_2)\log p(x_1,x_2).
\ee

Vibrational wavefunctions are given by the following expression
$\Psi_{n}(x,l) = \frac{e^{-\frac{x^2}{2 l^2}}}{\sqrt{2^n n!}} \frac{1}{\sqrt[4]{\pi l^2}} H_n(\frac{x}{l})$. 
\\In this expression $x$ gives coordinates of molecule, $l=\sqrt{\frac{\hbar}{m \omega}}$ -- normalization harmonic parameter depending on oscillating system, $H_n$ -- is the Hermite polynomial.

Squared norm of overlap integral of two vibrational wavefunctions for different electronic states gives the FCF-s -- probability of transition from first electronic state of vibrational level $m$ to the second electronic state of vibrational level $n$:

$$ P_{mn} (a) = \left| I_{mn}(a) \right| ^2 = \left|\int_{-\infty }^{\infty} \psi_m^*(x) \psi_n(x-a) dx \right| ^2, $$
where $a$ defines the shift between the minima between potential energy of two electronic states.

Calculation of overlap integral $I_{mn}(a)$ for the case $m=0$ with vibronic states with different harmonic parameters $l$: initial state with parameter $l=1$ and final state $l=l$ is linked with the following relation:

$I_{0n}(x,l)= \int_{-\infty}^{\infty} \psi_0^*(x,l=1) \psi_n(x-a,l) dx = \int_{-\infty}^{\infty} \psi_0^*(x+a,l=1) \psi_n(x,l) dx$.

Initial state is $\Psi_0(x+a,l=1) = \frac{e^{-\frac{(x+a)^2}{2}}}{\sqrt[4]{\pi}}$.

Final state is $\Psi_{n}(x,l) = \frac{e^{-\frac{x^2}{2 l^2}}}{\sqrt{2^n n!}} \frac{1}{\sqrt[4]{\pi l^2}} H_n(\frac{x}{l})$.

This gives us the following expression for FCF-s:

$$P_{0n}(a,l) = \left( \frac{l^2-1}{l^2+1} \right) ^n \frac{2 l}{l^2 + 1} \frac{e^{-\frac{a^2}{l^2 +1}}}{2^n n!} H_n^2 \left(-\frac{a l}{\sqrt{(l^2+1)(l^2-1)}} \right)$$

Expression for FCF-s can be interpreted as probability distribution of one random variable $P_{0n}(a,l) = P(n)$.

Employing the mentioned above approach \cite{Sei17} we can examine probabilistic properties of FCF-s.

Let us consider FCF-s as two-dimensional probability distribution. We can split a system of vibronic transitions into two subsystems:

$$P_{0n}(a,l) = P(n) = P(n(q_1, q_2)).$$

Because ground state has number $ n=0 $ expression in App. B has the following form:
$$n+1 = q_1+(q_2-1)Q_1,\qquad 1\leq q_1 \leq Q_1,\quad 1\leq q_2 \leq Q_2.$$

Bringing together three above expressions with $Q_1=2$ and $Q_2=\infty$ we obtain the following expression: 

$$P(n(q_1, q_2)) = \left( \frac{l^2-1}{2(l^2+1)} \right) ^{q_1+2 q_2-3} \frac{2 l}{l^2 + 1} \frac{e^{-\frac{a^2}{l^2 +1}}}{(q_1+2 q_2-3)!} H^2_{q_1+2 q_2-3}\left(-\frac{a l}{\sqrt{(l^2+1)(l^2-1)}} \right)$$

With this expression as probability distribution of two variables we can find two marginal probability distributions 

\be \mathcal{P}(q_1)=\sum_{q_2=1}^{\infty}P(n(q_1,q_2))
\ee
and

\be \Pi(q_2)=\sum_{q_1=1}^{2}P(n(q_1,q_2))
. \ee

The first probability distribution consist of two probabilities $\mathcal{P}(1)$ and $\mathcal{P}(2)$ which are the sums of odd and even terms of initial probability distribution.

\begin{eqnarray*}
	\mathcal{P}(1)=\sum_{n=0}^{\infty} P(2n)=\sum_{n=0}^{\infty} \left( \frac{l^2-1}{2(l^2+1)} \right) ^{2n} \frac{2 l}{l^2 + 1} \frac{e^{-\frac{a^2}{l^2 +1}}}{(2n)!} H^2_{2n}\left(-\frac{a l}{\sqrt{(l^2+1)(l^2-1)}} \right)
\end{eqnarray*}

\begin{eqnarray*} \label{margin1}
	\mathcal{P}(2)=\sum_{n=0}^{\infty} P(2n+1)=\sum_{n=0}^{\infty} \left( \frac{l^2-1}{2(l^2+1)} \right) ^{2n+1} \frac{2 l}{l^2 + 1} \frac{e^{-\frac{a^2}{l^2 +1}}}{(2n+1)!} H^2_{2n+1}\left(-\frac{a l}{\sqrt{(l^2+1)(l^2-1)}} \right)
\end{eqnarray*}

The second probability distribution consist of the following probabilities 

\begin{eqnarray*} \label{margin2}
	\Pi(k)= P(2k-2) + P(2k-1)=\\
	=\left( \frac{l^2-1}{2(l^2+1)} \right) ^{2k-2} \frac{2 l}{l^2 + 1} \frac{e^{-\frac{a^2}{l^2 +1}}}{(2k-2)!} H^2_{2k-2}\left(-\frac{a l}{\sqrt{(l^2+1)(l^2-1)}} \right) +\\
	+ \left( \frac{l^2-1}{2(l^2+1)} \right) ^{2k-1} \frac{2 l}{l^2 + 1} \frac{e^{-\frac{a^2}{l^2 +1}}}{(2k-1)!} H^2_{2k-1}\left(-\frac{a l}{\sqrt{(l^2+1)(l^2-1)}} \right)= \\
	= \left( \frac{l^2-1}{2(l^2+1)} \right) ^{2k-2} \frac{2 l}{l^2 + 1} \frac{e^{-\frac{a^2}{l^2 +1}}}{(2k-2)!} \left( H^2_{2k-2}\left(-\frac{a l}{\sqrt{(l^2+1)(l^2-1)}} \right) + \right.\\
	\left.+  \left( \frac{l^2-1}{2(l^2+1)} \right) \frac{1}{2k-1} H^2_{2k-1}\left(-\frac{a l}{\sqrt{(l^2+1)(l^2-1)}} \right) \right)=\Pi(k)
\end{eqnarray*}

Explicit expressions for marginal probability distributions allow us to rewrite subadditivity condition for FCF-s:

\begin{eqnarray}
-\sum_{k=1}^{2}\mathcal{P}(k)\log\mathcal{P}(k)
-\sum_{k=1}^{\infty}\Pi(k)\log\Pi(k) \geq
-\sum_{n=0}^{\infty} P(n)\log P(n).
\end{eqnarray}

Mutual information, which is given by the difference between the LHS and RHS of the previous equation is the function of shift $a$ and harmonic parameter $l$. We can plot this function

\begin{figure}
	\centering
	\subfloat[3D plot of the mutual information]{%
		\resizebox*{5cm}{!}{\includegraphics{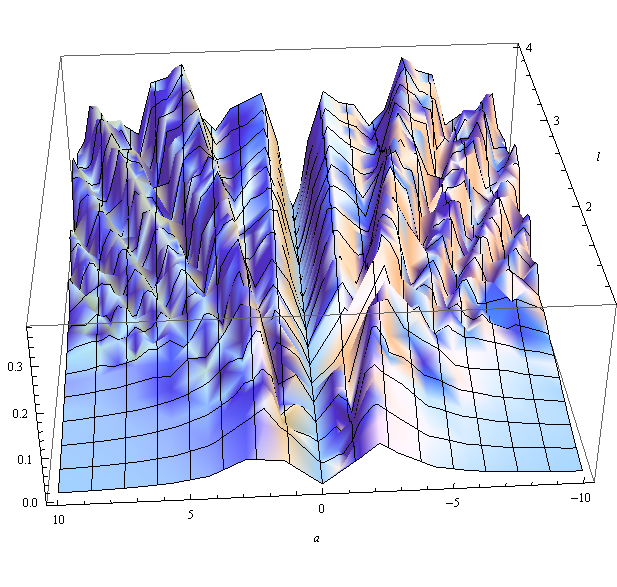}}}\hspace{5pt}
	\subfloat[grayscale plot of the mutual information, in which larger values are shown lighter]{%
		\resizebox*{5cm}{!}{\includegraphics{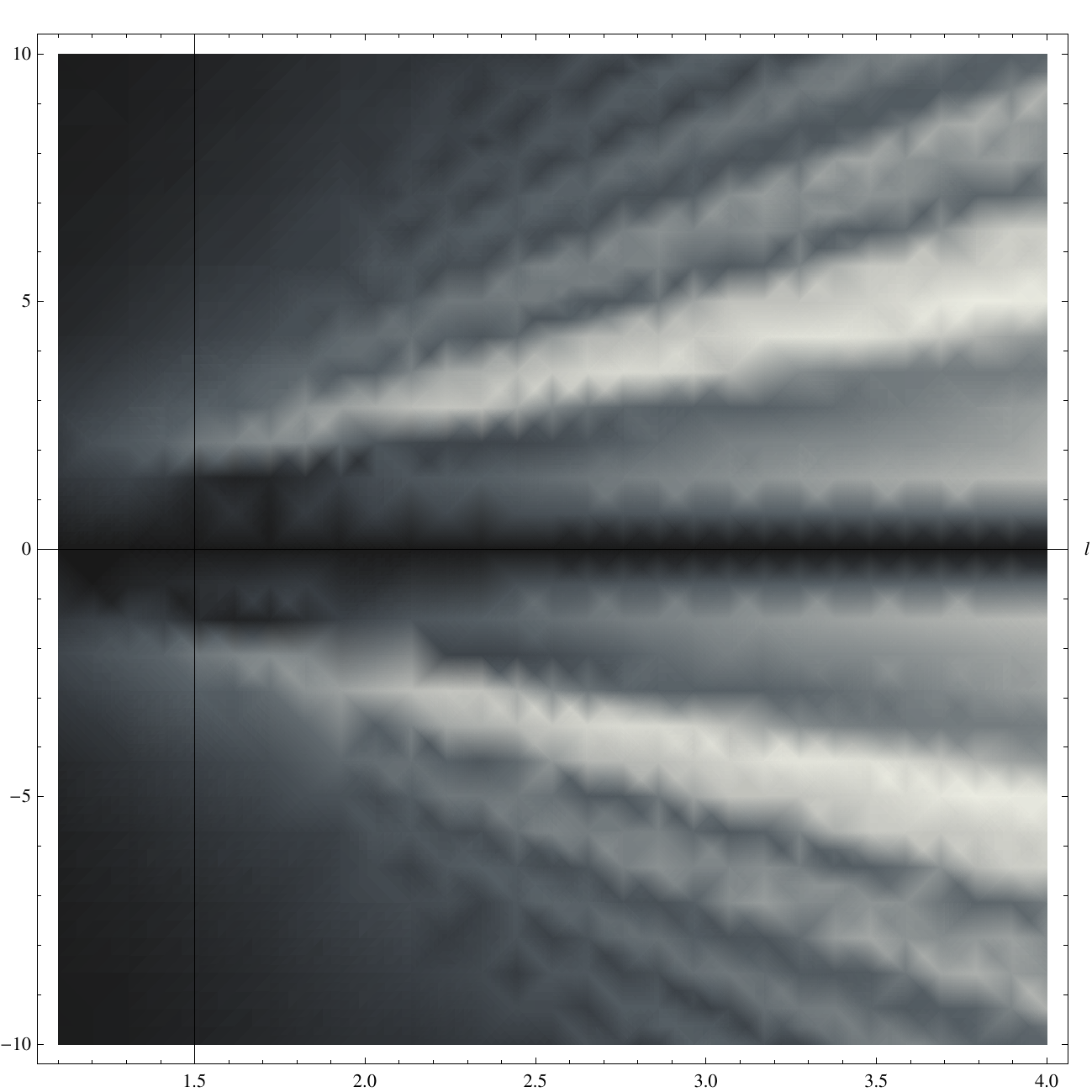}}}
	\caption{Mutual information as a function of shift $a$ and harmonic parameter $l$} \label{mutual}
\end{figure}

Obtained expression depicted on the Fig.\ref{mutual} shows correlations as a periodic function FCF-s depending on combinations of shift of the minimum of potential $a$ and harmonic parameter $l$ which is inversely proportional with the frequency of final oscillating system.

\section{Results and discussion}

To conclude we point out the main results of our work. Information-entropic characteristics of vibronic structure of electronic lines of diatomic molecules were introduced. The vibronic repetitions of electronic line intensities were described. New method to find the correlations between the intensities of vibronic repetitions of electronic lines for arbitrary potential including arbitrary unharmonic potentials was suggested. The notion of mutual information and entropy of artificial subsystems associated with the vibronic structure of the diatomic molecules was introduced. 

A periodicity of correlations was found in the case of harmonic potential for which the FCF-s for diatomic molecules are expressed in explicit form in terms of Hermite polynomials.

The developed approach to polyatomic molecules with the example of triatomic molecules will be extended in the following research.

The quantum correlations in vibronic structure of the molecules based on introduced in our work entropic-information analysis of FCF-s can be found also for polyatomic molecules. This generalization can be also used in quantum information physics and applications like quantum technologies.

\end{document}